\documentclass[twocolumn,showpacs,pra]{revtex4}
\usepackage{amssymb}
\usepackage{graphicx}
\usepackage{dcolumn}
\usepackage{bm}
\usepackage{amsmath}
\usepackage{epsfig}

\begin{document}

\title{Controlling Molecular Scattering by Laser-Induced Field-Free Alignment}

\author{E. Gershnabel}
\author{I.Sh. Averbukh}
\affiliation{Department of Chemical Physics, The Weizmann Institute
of Science, Rehovot 76100, ISRAEL}
\begin{abstract}

We consider deflection of  polarizable  molecules by inhomogeneous
optical fields, and analyze the role of molecular orientation and
rotation in the scattering process. It is shown that molecular
rotation induces spectacular rainbow-like features in the
distribution of the scattering angle. Moreover, by preshaping
molecular angular distribution with the help of short and strong
femtosecond laser pulses, one may efficiently control the scattering
process, manipulate the average deflection angle and its
distribution, and reduce substantially the angular dispersion of the
deflected molecules. We provide quantum  and classical treatment of the deflection process. The effects of strong deflecting field on the scattering of rotating molecules are considered by the means of the adiabatic invariants formalism. This new control scheme opens new ways for many applications
involving molecular focusing, guiding and trapping by optical and
static fields.

\end{abstract}

\maketitle

\section{Introduction} \label{Introduction}
Optical dipole forces acting on molecules in nonresonant laser fields
is a hot subject of many recent experimental studies
\cite{Deflection_general,Lens,Prism,Barker-new}. By controlling
molecular translational degrees of freedom with laser fields
\cite{Friedrich,Seideman,Friedrich1,Gordon,Fulton,Fulton1}, novel
elements of molecular optics can be realized, including molecular
lens \cite{Deflection_general,Lens} and molecular prism
\cite{Prism}.  The mechanism of molecular interaction with a nonuniform
laser field is rather clear: the field induces molecular
polarization, interacts with it, and deflects the molecules along
the intensity gradient.  As most molecules have anisotropic
polarizability, the deflecting force depends on the molecular
orientation with respect to the deflecting field. Previous studies
on optical molecular deflection have mostly considered randomly
oriented molecules, for which the deflection angle is somehow
dispersed around the mean value determined by the
orientation-averaged polarizability. The latter   becomes
intensity-dependent for strong enough fields due to the
field-induced modification of the molecular angular motion
\cite{Zon,Friedrich}. This adds a new ingredient for controlling
molecular trajectories  \cite{Friedrich,Seideman,Friedrich1,Gordon,Barker-new}, which is
important, but somehow limited because of using the same fields for
the deflection process and orientation control.

 In this work, we show that the deflection process can be significantly affected and controlled
 by  \textit{preshaping} molecular angular distribution \emph{before} the molecules
enter the interaction zone.  This can be done with the help of
numerous recent techniques for laser molecular alignment, which use
single or multiple short laser pulses (transform-limited, or shaped)
to align molecular axes along certain directions.  Short laser
pulses excite rotational wavepackets, which results in a
considerable transient molecular alignment after the laser pulse is
over, i.e. at field-free conditions (for  reviews on field-free
alignment, see, e.g. \cite{Stapelfeldt,Stapelfeldt1}). Field-free
alignment was observed both for small diatomic molecules as well as
for more complex molecules, for which full three-dimensional control
was realized \cite{3D1,3D2,3D3}.

We demonstrate that the average scattering angle of the deflected
molecules and its distribution may be dramatically modified by a
proper field-free prealignment. By separating the processes of the
angular shaping and the actual deflection, one gets a flexible tool
for tailoring molecular motion in inhomogeneous optical and static
fields.

The main principles of this new approach were briefly introduced in our
recent Letter \cite{ourPRL}. Here we present a much more elaborated analysis
of the control mechanisms, including also  a detailed comparison between the quantum and classical
aspects of the problem, and discussion of the strong field effects in molecular scattering.

In Sec. \ref{Deflection_heuristic} we present the deflection scheme,
as well as heuristic classical discussion on the anticipated role of
molecular rotation on the deflection process (both for thermal and
prealigned molecules). In Sec \ref{Quantum_Treatment} we verify
these predictions by the means of the quantum treatment of the
problem in the limit of the relatively weak deflecting field (that
does not disturb significantly the rotational motion). The strength
of the prealigning field is not restricted here. Full classical
treatment of the molecular deflection at such conditions (including
thermal effects) is given in Sec. \ref{Classical_Treatment_weak},
where we find a good correspondence between the classical and
quantum calculations. Motivated by this agreement, we provide in
Sec. \ref{classical_treatment_strong} a full classical analysis of
the molecular scattering by strong deflecting field using the adiabatic invariants
formalism. Finally we summarize our results in Sec.
\ref{Discussion_and_conclusions}.

\section{Deflection of field-free aligned molecules} \label{Deflection_heuristic}

\begin{figure}[htb]
\begin{center}
\includegraphics[width=8cm]{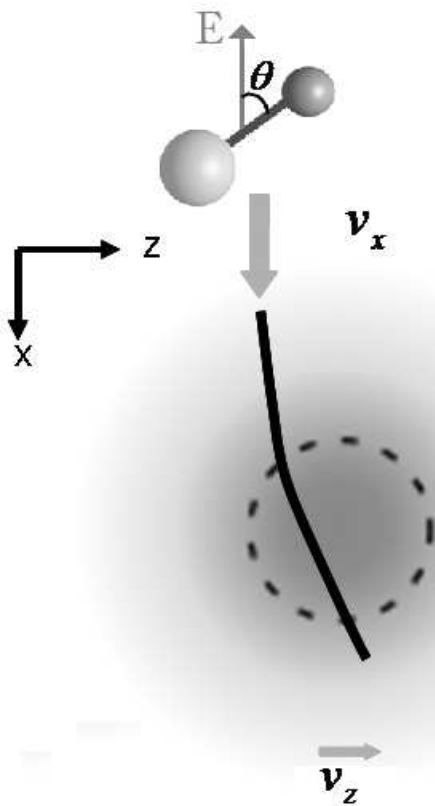}
\end{center}
\caption{The deflection scheme. A polarized (in the $z$ direction)
laser field propagates toward the plane of the paper ($y$
direction). The linear molecules, initially moving along the $x$
direction (with velocity $v_x$), are deflected by the potential
gradient (deflection velocity $v_z$).} \label{DeflectionScheme}
\end{figure}

Although our arguments are rather general, we follow for certainty a
deflection scheme that brings to mind the experiment by Stapelfeldt $et$
$al$ \cite{Deflection_general} who used a strong IR laser to deflect
a $CS_2$ molecular beam, and then addressed a portion of the
deflected molecules (at a preselected place and time) by an
additional short and narrow ionizing pulse. Consider  deflection (in
$z$ direction) of a linear molecule moving in $x$ direction  with
velocity $v_x$ and interacting with a focused nonresonant laser beam
that propagates along the $y$ axis (Fig. \ref{DeflectionScheme}).

The spatial profile of the laser electric field in the $xz$-plane
is:
\begin{equation}
E=E_0\exp [-(x^2+z^2)/\omega_0^2 ]\exp [-2\ln2t^2/\tau^2
].\label{E_deflect_field}
\end{equation}
The interaction potential of a linear molecule in the laser field is
given by:
\begin{equation}
U=-\frac{1}{4}E^2\left(\alpha_\parallel\cos^2\theta+\alpha_\perp\sin^2\theta\right),\label{U_general_eq}
\end{equation}
where $E$ is defined in Eq. \ref{E_deflect_field}, and
$\alpha_\parallel$ and $\alpha_\perp$ are the components of the
molecular polarizability along the molecular axis, and perpendicular
to it, respectively. Here $\theta$ is the angle between the electric
field polarization direction (along the laboratory $z$ axis) and the
molecular axis. A molecule initially moving along the $x$ direction
will acquire a velocity component $v_z$ along $z$-direction. We
consider the perturbation regime corresponding to a small deflection
angle, $\gamma\thickapprox v_z/v_x$. We substitute $x=v_x t$,  and
consider $z$ as a fixed impact parameter. The deflection velocity is
given by:

\begin{equation}\label{Velocity_Deflection}
v_z = \frac{1}{M}\int_{-\infty}^{\infty}F_z dt
=-\frac{1}{M}\int_{-\infty}^{\infty}\left(\overrightarrow{\nabla}U\right)_z
dt.
\end{equation}

Here $M$ is the mass of the molecules, and $F_z$ is the deflecting
force. The time-dependence of the force $F_z$ (and potential $U$) in
Eq.(\ref{Velocity_Deflection}) comes from three sources: pulse
envelope, projectile motion of the molecule through the laser focal
area, and time variation of the angle $\theta$ due to molecular
rotation. For simplicity, we start with the case of the relatively
weak deflecting field that does not affect significantly the
rotational motion. Such approximation is justified, say for $CS_2$
molecules with the rotational temperature $T=5K$, which are subject
to the deflecting field of $3\cdot10^9 W/cm^2$. The corresponding
alignment potential
$U\approx-\frac{1}{4}\left(\alpha_\parallel-\alpha_\perp\right)E_0^2\approx
0.04\ meV$  is an order of magnitude smaller than the thermal energy
$k_BT$, where $k_B$ is Boltzmann's constant. This assumption is even
more valid if the molecules were additionally subject to the
aligning pulses prior to deflection. The case of a strong deflecting
field will be considered later in
Sec.\ref{classical_treatment_strong}.

  Since the rotational time scale is the shortest one in the problem,
 we average  the force over the fast rotation, and arrive at the following expression
 for the deflection angle, $\gamma = v_z/v_x$:
\begin{equation}\label{Deflection Angle}
\gamma = \gamma_0 \ \left[\alpha_{||}{\cal A} +\alpha_{\bot} (1
-{\cal A} )\right]/\overline{\alpha}
\end{equation}
Here $\overline{\alpha}=1/3 \alpha_{||}+2/3 \alpha_{\bot}$ is the
orientation-averaged molecular polarizability, and ${\cal
A}=\overline{\cos^2\theta}$ denotes the time-averaged value of
$\cos^2\theta$. This quantity depends on the relative orientation of
the vector of angular momentum and the polarization of the
deflecting field. It is different for different molecules of the
incident ensemble, which leads to the randomization of the
deflection process. The constant $\gamma_0$ presents the average
deflection angle for an isotropic molecular ensemble:
\begin{eqnarray}\label{Average_Deflection Angle}
\gamma_0 &=&\frac{\overline{\alpha}E_0^2}{4Mv_x^2}\left(\frac{-4z}{\omega_0}\right) \nonumber\\
&\times&\sqrt{\frac{\pi}{2}}\left(1+\frac{2\omega_0^2\ln2}{\tau^2v_{x}^2}\right)^{-1/2}\exp\left(-\frac{2z^2}{\omega_0^2}\right)
\end{eqnarray}
We provide below some heuristic classical arguments on the
anticipated statistical properties  of ${\cal A}$ and  $\gamma$
(both for thermal and prealigned molecules).

\begin{figure}[htb]
\begin{center}
\includegraphics[width=2cm]{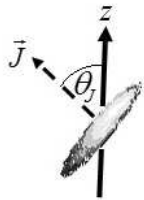}
\end{center}
\caption{A molecule rotates with the angular momentum $\vec{J}$
forming an angle $\theta_J$ with the laboratory $z$ axis.}
\label{SimpleModel}
\end{figure}

Consider a linear molecule that rotates freely in a plane that is
perpendicular to the vector $\overrightarrow{J}$ of the angular
momentum (see Fig.(\ref{SimpleModel})).
\\

The projection of the molecular axis on the vertical $z$-direction
is given by:
\begin{equation}
\cos \theta(t)=  \cos(\omega t)
\sin\theta_J,\label{Simple_COS_model}
\end{equation}
where $\theta_J$ is the angle between $\vec{J}$ and $z$-axis, and
$\omega$ is the angular frequency of molecular rotation. Averaging
over time, one arrives at:
\begin{equation}
{\cal A}=\overline{\cos^2\theta}=\frac{1}{2}
\sin^2\theta_J.\label{Simple_COS_avg_time}
\end{equation}
In a \emph{thermal} ensemble, vector $\vec{J}$ is randomly oriented
in space, with  isotropic angular distribution density $1/2 \sin(\theta_J
)$.  The mean value of the deflection angle is $\langle
\gamma \rangle=\gamma_0$.  Eq.(\ref{Simple_COS_avg_time}) allows us
to obtain the distribution function, $f({\cal A})$ for ${\cal A}$
(and the related deflection angle) from the known isotropic
distribution for $\theta_J$. Since the inverse function
$\theta_J({\cal A})$ is multivalued, one obtains
\begin{equation}
f({\cal A})=\sum_{i=1}^2 \frac{1}{2} \sin\theta_J^{(i)}{\left|
\frac{d{\cal
A}}{d\theta_J^{(i)}}\right|}^{-1}=\frac{1}{\sqrt{1-2{\cal
A}}},\label{Gamma_Dist}
\end{equation}
where we summed over the two branches of $\theta_J({\cal A})$. This
formula predicts an \emph{unimodal  rainbow} singularity in the
distribution of the scattering angles at the maximal value
$\gamma=\gamma_0 (\alpha_{||} +\alpha_{\bot})/2\overline{\alpha}$ \
(for ${\cal A}=1/2$), and a flat step near the minimal one
$\gamma=\gamma_0 \alpha_{\bot}/\overline{\alpha}$ \ (for ${\cal
A}=0$).
\\

Assume now that the molecules are prealigned before entering the
deflection zone by a strong and short  laser pulse that is polarized
\emph{perpendicular} to the polarization direction of the deflecting
field (e.g., in $x$-direction). Such a pulse forces the molecules to
rotate preferentially in the planes containing the $x$-axis. As a
result, the vector $\vec{J}$ of the angular momentum is confined to
the $yz$-plane, and angle $\theta_J$ becomes uniformly distributed
in the interval $[0,\pi]$ with probability density $1/\pi$.
 The corresponding probability distribution for ${\cal A}$  takes the form
\begin{equation}
f({\cal A})=\frac{\sqrt{2}}{\pi}\frac{1}{\sqrt{{{\cal A}(1-2{\cal
A}})}} \label{Gamma_Dist1}
\end{equation}
In contrast to Eq.(\ref{Gamma_Dist}), Eq.(\ref{Gamma_Dist1})
suggests a \emph{bimodal rainbow} in the distribution of deflection
angles, with singularities both at the minimal and the maximal
angles. Finally, we proceed to the most interesting case when the
molecules are prealigned by a short strong laser pulse that is
polarized \emph{parallel} to the direction of the deflecting field.
After excitation by such a pulse, the vector of the angular momentum
of the molecules is preferentially confined to the $xy$-plane, and
the angle $\theta_J$ takes a well defined value of
$\theta_J\approx\pi/2$ which corresponds to ${\cal A}=1/2$. In this way, the
molecules experience the maximally possible time-averaged
deflecting force which is the same for all the
particles of the ensemble. As a result, the dispersion of the
scattering angles is  reduced dramatically. The distribution of the
deflection angle $\gamma$
  transforms to a narrow peak (asymptotically - a
$\delta$-function) near the maximal value, $\gamma=\gamma_0
(\alpha_{||} +\alpha_{\bot})/2\overline{\alpha}$.

\section{Quantum Treatment} \label{Quantum_Treatment}

\begin{figure}[htb]
\begin{center}
\includegraphics[width=8cm]{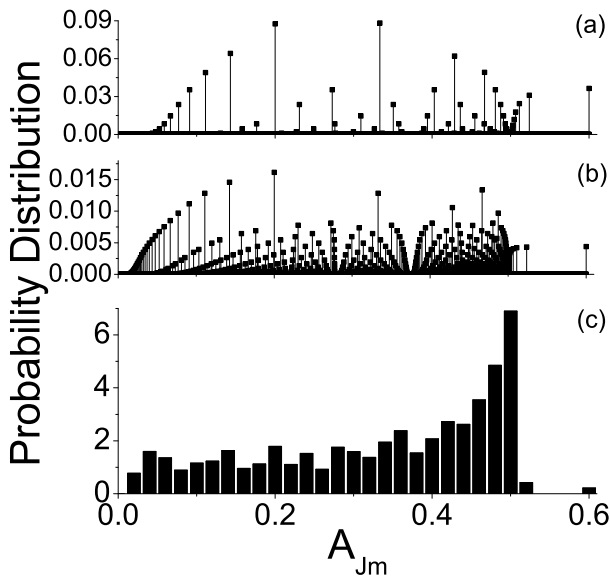}
\end{center}
\caption{Quantum distribution of ${\cal A}_{J,m}$ in the thermal
case. Panels (a) and (b) correspond to   $J_T=5$ and $J_T=15$,
respectively. Histogram in panel (c) shows a coarse-grained
version of the distribution in panel (b).}\label{Quant thermal}
\end{figure}

For a more quantitative treatment, involving analysis of the
relative role of the quantum and thermal effects on one hand, and
the strength of the prealigning pulses on the other hand, we
consider quantum-mechanically the deflection of a linear molecule
described by the Hamiltonian:
\begin{equation}
{\cal H}= \hat{J}^2/(2I).\label{Quantum_Hamiltonian}
\end{equation}
Here $\hat{J}$ is operator of angular momentum, and $I$ is the
moment of inertia, which is related to the molecular rotational
constant, $B=\hbar/(4\pi Ic)$  ($c$ is speed of light). Assuming
again that the deflecting field is too weak to modify molecular
alignment, we consider scattering in different $|J,m\rangle$ states
independently. The deflection angle is given by Eq.(\ref{Deflection
Angle}), in which ${\cal A}$ is replaced by
\begin{equation}
{\cal A}_{J,m}=\langle J,m|\cos^2\theta |J,m\rangle
=\frac{1}{3}+\frac{2}{3}\frac{J(J+1)-3m^2}{(2 J + 3) (2 J - 1)}.
\label{AJm}
\end{equation}
In the quantum case, the continuous distribution of the angles
$\gamma$ is replaced by a set of discrete lines, each of them
weighted by the population of the state $|J,m\rangle$. Fig.
\ref{Quant thermal} shows the distribution of ${\cal A}_{J,m}$ in
the thermal case for various values of the dimensionless parameter
$J_T=\sqrt{k_B T/(hBc)}$ that represents the typical "thermal" value
of $J$ (for $J_T\geq 1$). For $CS_2$ molecules, the values of $J_T=5,15$ correspond
to $T=3.9K$ and $T=35K$, respectively.

The distribution of discrete values of ${\cal A}_{J,m}$ demonstrates
a non-trivial pattern. In particular, the values exceeding the classical limit $0.5$ correspond to the states $|J,m=0\rangle$ (see Eq.(\ref{AJm})), and they rapidly approach that limit as $J$ grows. After the coarse-grained averaging, however, the distribution shows the expected unimodal
rainbow feature (see Eq.(\ref{Gamma_Dist})) for large enough $J_T$.

If the molecules are subject to a strong femtosecond prealigning
pulse, the corresponding interaction potential is given by Eq.
(\ref{U_general_eq}), in which $E$ is replaced by the envelope
$\epsilon$ of the femtosecond pulse.  If the pulse is short compared
to the typical periods of molecular rotation, it may be considered
as a delta-pulse. In the impulsive approximation, one obtains the
following relation between the angular wavefunction before and after
the pulse applied at $t=0$ (see e.g. \cite{Gershnabel}, and
references therein):
\begin{equation}
\Psi(t=0^+)=\exp(iP\cos^2\theta)\Psi(t=0^-),\label{before_after}
\end{equation}
where the kick strength, $P$ is given by:
\begin{equation}
P=\left(1/4\hbar\right)\cdot
(\alpha_{||}-\alpha_{\bot})\int_{-\infty}^{\infty}\epsilon^2(t)dt.\label{Kick_P}
\end{equation}
Here we assumed the vertical polarization (along $z$-axis) of the
pulse. Physically, the dimensionless kick strength, $P$ equals to
the typical amount of angular momentum (in the units of $\hbar$)
supplied by the pulse to the molecule. For example, in the case of $CS_2$ molecules,
the values of $P=5, 25$ correspond to the excitation by $0.5 ps$ (FWHM) laser pulses with the maximal intensity of $4.6\cdot10^{11} W/cm^2$ and  $2.3\cdot10^{12} W/cm^2$, respectively. For the vertical polarization
of the laser field, $m$ is a conserved quantum number. This allows
us to consider the excitation of the states with different initial
$m$ values separately. In order to find $\Psi(t=0^+)$ for any
initial state, we introduce an artificial parameter $\xi$ that will
be assigned the value $\xi=1$ at the end of the calculations, and
define
\begin{equation}
\Psi_{\xi}=\exp\left[(iP\cos^2\theta)\xi\right]\Psi(t=0^-)
=\sum_{J}c_J(\xi)|J,m\rangle.\label{xi_relation}
\end{equation}
By differentiating both sides of Eq.(\ref{xi_relation}) with respect
to $\xi$, we obtain the following set of differential equations for
the coefficients $c_J$:
\begin{equation}
\dot{c}_{J'}=iP\sum_J c_J\langle
J',m|\cos^2\theta|J,m\rangle,\label{Differential equations}
\end{equation}
where $\dot{c}= dc / d\xi $. The diagonal matrix elements in
Eq.(\ref{Differential equations}) are given by Eq.(\ref{AJm}), the
off-diagonal ones can be found using recurrence relations for the
spherical harmonics \cite{Arfken}. Since $\Psi_{\xi=0}=\Psi(t=0^-)$
and $\Psi_{\xi=1}=\Psi(t=0^+)$ (see Eq.(\ref{xi_relation})), we
solve numerically this set of equations from $\xi=0$ to $\xi=1$, and
find $\Psi(t=0^+)$. In order to consider the effect of the
field-free alignment at thermal conditions, we repeated this
procedure for every initial $|J_0,m_0\rangle$ state. To find the
modified population of the $|J,m\rangle$ states, the corresponding
contributions from different initial states were summed together
weighted with the Boltzmann's statistical factors:
\begin{eqnarray}
f({\cal
A}_{J,m_0})&=&\sum_{J_0,\bar{J}}\frac{\exp(-E^{J_0}/K_BT)}{Q_{rot}}
\nonumber\\ &\times& |c_{\bar{J}}|^2\delta ({\cal A}_{J,m_0},{\cal
A}_{\bar{J},m_0}),\label{QuantDistribution}
\end{eqnarray}
where $c_J$ are the coefficients (from Eq. \ref{Differential
equations}) of the wave packet that was excited from the initial
state $|J_0,m_0\rangle$; $\delta$ is the Kronecker delta, and
$Q_{rot}$ is the rotational partition function.  It worth mentioning
that different combinations of $J$ and $m$ may correspond to the
same value of ${\cal A}_{J,m}$, which necessitates the presence of
the Kronecker delta in Eq. (\ref{QuantDistribution}).
For symmetric molecules, statistical spin factor should be taken
into account. For example, for $CS_2$ molecules in the ground
electronic and vibrational state, only even $J$ values are allowed
due to the permutation symmetry for the exchange of two Bosonic
Sulfur atoms (that have nuclear spin $0$).

In the case of an aligning pulse in the $x$ direction, the operator
in Eq.(\ref{before_after}) becomes:
\begin{equation}
\Psi(t=0^+)=\exp(iP\cos^2\phi\sin^2\theta)\Psi(t=0^-),\label{before_after_perpendicular}
\end{equation}
and a similar procedure as described above is used to find the
deflection distribution. One should pay attention that $m$ is no
longer a conserved quantum number for a pulse kicking in the $x$
direction.

\begin{figure}[htb]
\begin{center}
\includegraphics[width=8cm]{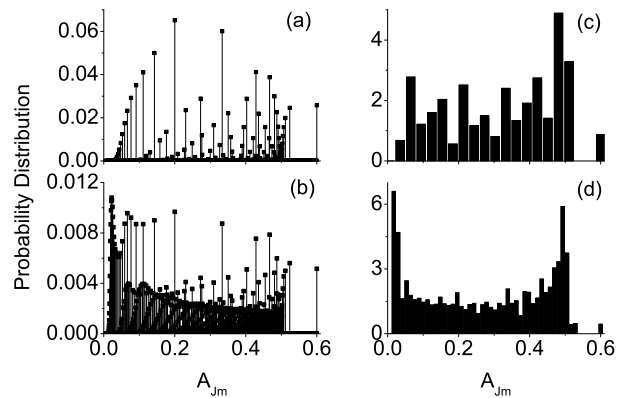}
\end{center}
\caption{Distribution of ${\cal A}_{J,m}$ for molecules prealigned
with the help of a short laser pulse polarized in the $x$ direction.
The left column (a-b) presents directly the ${\cal A}_{J,m}$ values,
while the right column (c-d) shows the corresponding coarse-grained
histograms (as in  Fig. \ref{Quant thermal}c). Panels (a) and (c)
are calculated for $J_T=5$ and $P=5$; (b) and (d) are for $J_T=5$
and $P=25$.} \label{DistributionT1P10}
\end{figure}
\begin{figure}[htb]
\begin{center}
\includegraphics[width=8cm]{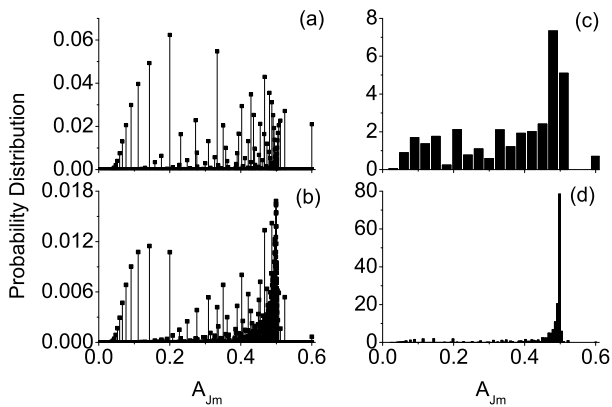}
\end{center}
\caption{Distribution of ${\cal A}_{J,m}$ for molecules prealigned
in the $z$ direction. The left column (a-b) presents directly the
${\cal A}_{J,m}$ values, while the right column (c-d) shows the
corresponding coarse-grained histograms. Panels (a) and (c) are
calculated for $J_T=5$ and $P=5$; (b) and (d) are for $J_T=5$ and
$P=25$.} \label{QuantDistP25T5}
\end{figure}
Using this technique, we considered deflection of initially thermal
molecules that were prealigned with the help of short pulses
polarized in $x$ and $z$ directions (Figs. \ref{DistributionT1P10}
and \ref{QuantDistP25T5}, respectively). In the case of the
alignment \emph{perpendicular} to the deflecting field (Fig.
\ref{DistributionT1P10}), the coarse-grained distribution of ${\cal
A}_{J,m}$ (and that of the deflection angle) exhibits the
\emph{bimodal rainbow} shape, Eq.(\ref{Gamma_Dist1}) for strong
enough kicks ($P\gg 1$ and $P\gg J_T$). Finally, and most
importantly, prealignment in the direction \emph{parallel} to the
deflecting field allows for almost complete removal of the
rotational broadening. A considerable narrowing of the distribution
can be seen when comparing Fig. \ref{Quant thermal}a and Figs.
\ref{QuantDistP25T5}b and \ref{QuantDistP25T5}d. The conditions required for the considerable narrowing
shown at Fig. \ref{QuantDistP25T5}d correspond to the maximal degree of field-free pre-alighment  $\langle \cos^2 \theta \rangle_{max} = 0.7$. This can be readily achieved with the current experimental technology, even at room temperature \cite{eightpulses}.

\section{Classical Treatment: weak deflecting field} \label{Classical_Treatment_weak}

Consider a classical rigid rotor (linear molecule) described by a
Lagrangian
\begin{equation}
L=\frac{1}{2}I(\dot{\theta}^2+\dot{\phi}^2\sin^2 \theta
),\label{Lagrangian}
\end{equation}
where $\theta$ and $\phi$ are Euler angles, and $I$ is the moment of
inertia. The canonical momentum for the $\phi$ angle
\begin{equation}
P_{\phi}= I \dot{\phi} \sin^2 \theta \label{Pphi}
\end{equation}
is a constant of motion as $\phi$ is a cyclic coordinate. The
canonical momentum $P_{\theta}$ is given by
\begin{equation}
P_{\theta}= I \dot{\theta}  \label{Ptheta}
\end{equation}
The Euler-Lagrange equation for the $\theta$ variable is
\begin{equation}
\frac{d}{dt}\frac{\partial L}{\partial\dot{\theta}}-\frac{\partial
L}{\partial\theta}=0, \label{EL}
\end{equation}
which leads to
\begin{equation}
\ddot{\theta} =
\frac{P_{\phi}^2}{I^2}\frac{\cos\theta}{\sin^3\theta}\label{thetaeq}
\end{equation}
When considering a thermal ensemble of molecules, it is convenient
to switch to dimensionless variables, in which the canonical momenta
are measured in the units of $p_{th}=I\omega_{th}$,  with
$\omega_{th}=\sqrt{k_B T/I}$, where $T$ is the temperature
\cite{Gershnabel}. By setting $P_{\phi}'=P_{\phi}/p_{th}$,
$P_{\theta}'=P_{\theta}/p_{th}$, and $t'=\omega_{th}t$, one gets the
following solution of Eq.(\ref{thetaeq}):
\begin{eqnarray}\label{COS_Classical}
\cos\theta(t')&=&\frac{1}{2}
\left(1-\frac{P_{\theta}'(0)}{\omega}\right)\cos(\theta(0)-\omega
t')\nonumber\\
&+&\frac{1}{2}\left(1+\frac{P_{\theta}'(0)}{\omega}\right)\cos(\theta(0)+\omega
t'),
\end{eqnarray}
where
\begin{equation}
\omega=\left(P_{\theta}'^2 (0)+\frac{P_{\phi}'^2
(0)}{\sin^2\theta(0)}\right)^{1/2}. \label{oomega}
\end{equation}
As in Sec \ref{Quantum_Treatment}, if at $t=0$ the molecules are
subject to a femtosecond aligning pulse polarized in $z$-direction,
the corresponding interaction potential is given by Eq.
(\ref{U_general_eq}), in which $E$ is replaced by the envelope
$\epsilon$ of the femtosecond pulse. We assume again that such a
pulse is short compared to the rotational period of the molecules,
and consider it as a delta-pulse. The rotational dynamics of the
laser-kicked molecules is then described by Eq.
(\ref{COS_Classical}), in which $P_{\theta}' (0)$ is replaced by
\begin{equation}
 P_{\theta}'(0)\rightarrow P_{\theta}'(0)-P_s'\sin(2\theta(0))\label{Parallel_classical}.
\end{equation}
Here $P_s '=P\hbar /\sqrt{k_BTI}$ is properly normalized kick
strength \cite{Gershnabel} with $P$ given by Eq. (\ref{Kick_P}).

In the case of an aligning pulse in the $x$ direction, both
$P_{\theta}'(0)$ and $P_{\phi}'(0)$ are replaced by:
\begin{eqnarray}
P_{\theta}'(0)&\rightarrow& P_{\theta}'(0)+P_s'\cos^2\phi(0)\sin(2\theta(0))\nonumber\\
P_{\phi}'(0)&\rightarrow&
P_{\phi}'(0)-P_s'\sin^2(\theta(0))\sin(2\phi(0))\label{PerpendicularPulses}
\end{eqnarray}

Averaging $\cos^2\theta(t')$ over time, we obtain:
\begin{eqnarray}
{\cal A} &=&\overline{\cos^2\theta} =
\frac{1}{4}\left[1+\left(\frac{P_{\theta}'(0)}{\omega}\right)^2\right]\nonumber\\
&+&\frac{1}{4}\left[1-\left(\frac{P_{\theta}'(0)}{\omega}\right)^2\right]\cos(2\theta(0)),
\label{COS2_Classical}
\end{eqnarray}
and the probability distribution of the time-averaged alignment
factor ${\cal A}$ can be obtained by:
\begin{eqnarray}
f({\cal A})&=&\int\int\int\int d\theta(0) d\phi(0) dP_{\theta}'(0)
dP_{\phi}'(0) \delta({\cal A}-\overline{\cos^2\theta})
\nonumber\\
&\times&F[\theta(0),\phi(0),P_{\theta}'(0),P_{\phi}'(0)]\label{Accurate_Dist}
\end{eqnarray}

where
\begin{equation}
F=\frac{1}{8\pi^2}\exp\left[-\frac{1}{2}\left(P_\theta'^2+\frac{P_{\phi}'^2}{\sin^2\theta}\right)\right]
\label{BoltDist}
\end{equation}
is the thermal distribution function.

\begin{figure}[htb]
\begin{center}
\includegraphics[width=8cm]{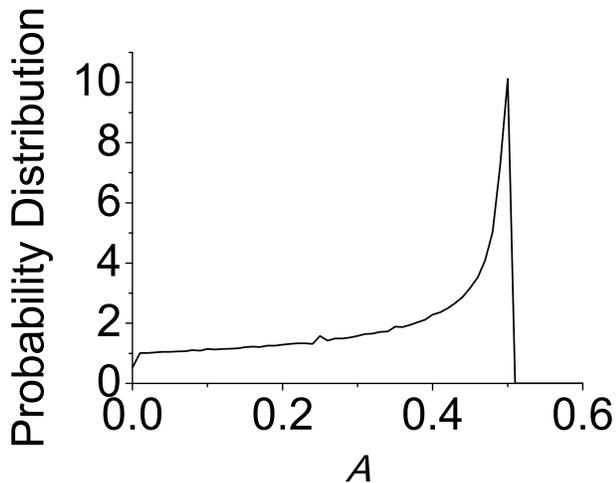}
\end{center}
\caption{Classical distribution of ${\cal A}$ for $J_T=15$. No
prealignment is assumed ($P=0$). One can observe a rainbow-like
feature at the right edge of the distribution, and a flat step at
the left edge.} \label{ClassicalDistJ15P0}
\end{figure}

The probability distribution of ${\cal A}$ in the thermal case is
plotted in Fig. \ref{ClassicalDistJ15P0}. Its shape is well
described by Eq. \ref{Gamma_Dist},  and it is in good agreement with
the quantum result of Fig. \ref{Quant thermal}c.

\begin{figure}[htb]
\begin{center}
\includegraphics[width=8cm]{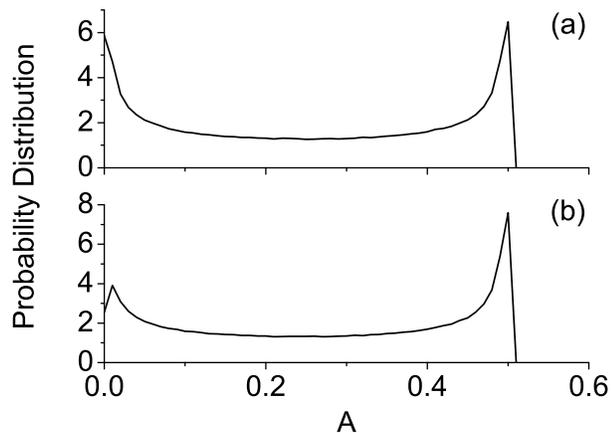}
\end{center}
\caption{Classical distribution of ${\cal A}$ for (a) $J_T=0.5$ and
(b) $J_T=5$ after the molecules were prealigned by a laser pulse
($P=25$) in the $x$ direction (i.e. perpendicular to the deflecting
field). Figure (a) corresponds to the case $P\gg J_T$, and it is in
good agreement with the analytical result, Eq.
(\ref{Gamma_Dist1}).}\label{PerpendicularDist}
\end{figure}

\begin{figure}[htb]
\begin{center}
\includegraphics[width=8cm]{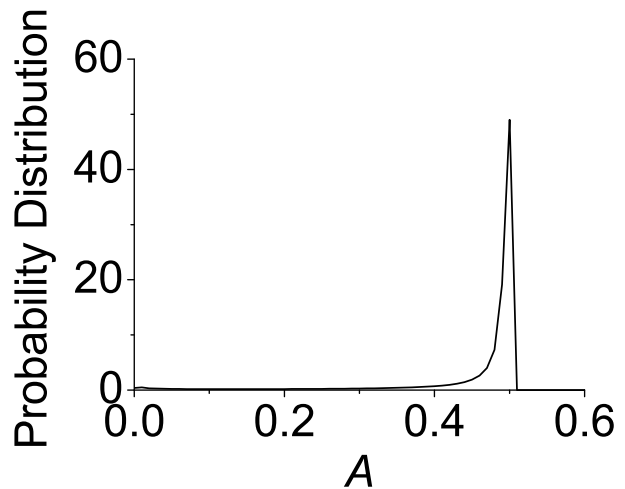}
\end{center}
\caption{Narrow classical distribution of $\cal A$
results from prealignment by  means of a laser pulse polarized
parallel to the deflecting field ($P=25$ and $J_T=5$).}
\label{PzClassicalDist}
\end{figure}

Figs. \ref{PerpendicularDist} and \ref{PzClassicalDist} show the
distribution of ${\cal A}$ value for molecules that were prealigned
in the direction perpendicular and parallel to the deflection field,
respectively.

In the case of perpendicular prealignment by a sufficiently strong
kick ($P\gg J_T$), the  distribution shown in Fig.
\ref{PerpendicularDist}a demonstrates the bimodal rainbow shape
predicted by Eq. (\ref{Gamma_Dist1}). Figure \ref{PerpendicularDist}b is
 similar to the corresponding quantum histogram in Fig.
\ref{DistributionT1P10}d.

In the case of parallel prealignment, the predicted narrow distribution is seen in Fig.
\ref{PzClassicalDist}. In what follows, we provide an asymptotic estimate of the width,
$\Delta{\cal A}$ of this distribution, and the mean value $<{\cal A}>$ of ${\cal A}$
in the limit of $P/J_T\gg 1$.

For strong enough kicks, Eq. (\ref{COS2_Classical}) shows that ${\cal A}$ approaches the value of
$\frac{1}{2}$, unless $\theta(0)$  is close to $\frac{\pi}{2}$. Therefore, we
define
\begin{equation}
\delta {\cal A}\equiv{\cal A}-\frac{1}{2}. \label{A_bar}
\end{equation}
that is different from zero only for small values of  $\beta = \theta(0) - \frac{\pi}{2}$.
For $\theta(0)\approx \frac{\pi}{2}$, Eqs. (\ref{Parallel_classical}),
(\ref{COS2_Classical}) and (\ref{A_bar}) yield:

\begin{equation}
\delta {\cal A}\approx\frac{1}{2}\left[
-\frac{P_{\phi}'(0)^2}{\left(P_{\theta}'(0)+2\beta P_s'
\right)^2+P_{\phi}'(0)^2} \right]. \label{A_bar_approx}
\end{equation}

The thermal averaging provides:

\begin{eqnarray}
\langle \delta {\cal A}\rangle
&\approx&-\frac{1}{2}\int_{-\infty}^{\infty}
dP_{\theta}'(0)dP_{\phi}'(0)\nonumber\\ &\times&
\int_{-\xi}^{\xi}d\beta
\frac{P_{\phi}'(0)^2}{\left(P_{\theta}'(0)+ 2\beta P_s'
\right)^2+P_{\phi}'(0)^2}\nonumber\\
&\times&\frac{\exp\left\{-\frac{1}{2}\left
[P_{\theta}'(0)^2+P_{\phi}'(0)^2 \right]\right\}}{Q_{rot}},
\label{Avg_classical_A}
\end{eqnarray}
where $Q_{rot}$ is the rotational partition function, and $[-\xi,+\xi]$
is the interval of the $\beta$ values for which the approximation (\ref{A_bar_approx}) is valid . We
continue manipulating the expression by introducing $\gamma = 2\beta P_s'$:

\begin{eqnarray}
\langle \delta {\cal A}\rangle
&\approx&-\frac{1}{4P_s'}\int_{-\infty}^{\infty}
dP_{\theta}'(0)dP_{\phi}'(0)\nonumber\\ &\times&
\int_{-2P_s'\xi}^{2P_s'\xi}d\gamma
\frac{P_{\phi}'(0)^2}{\left(P_{\theta}'(0)+ \gamma
\right)^2+P_{\phi}'(0)^2}\nonumber\\
&\times&\frac{\exp\left\{-\frac{1}{2}\left
[P_{\theta}'(0)^2+P_{\phi}'(0)^2 \right]\right\}}{Q_{rot}}.
\label{Avg_classical_A_simplified}
\end{eqnarray}
In the limit of $P_s' \rightarrow \infty$, the leading term in the asymptotic expansion of $\langle \delta {\cal A}\rangle$ can be obtained by expanding the limits of the internal integration $\pm 2P_s'\xi$ to $\pm\infty$ (as the integrand vanishes for large values of $P_s'\xi$):

\begin{eqnarray}
\langle \delta {\cal A}\rangle
&\approx&-\frac{1}{4P_s'}\int_{-\infty}^{\infty}
dP_{\theta}'(0)dP_{\phi}'(0)|P_{\phi}'(0)|\nonumber\\ &\times&
\int_{-\infty}^{\infty}d\gamma
\frac{1}{\gamma^2+1}\nonumber\\
&\times&\frac{\exp\left\{-\frac{1}{2}\left
[P_{\theta}'(0)^2+P_{\phi}'(0)^2
\right]\right\}}{Q_{rot}}\nonumber\\&=&-\frac{1}{P_s'}\cdot
\sqrt{\frac{\pi}{32}}. \label{Avg_classical_A_simplified2}
\end{eqnarray}
Recalling Eq. (\ref{A_bar}),  we  conclude that:
\begin{equation}
\langle {\cal A} \rangle \approx \frac{1}{2}-\frac{1}{P_s'}\cdot
\sqrt{\frac{\pi}{32}}. \label{Final_average_A}
\end{equation}

In order to estimate the width of the ${\cal A}$ distribution, we need to
consider the dispersion, and accordingly the average value of ${\cal
A}^2$. Following the same procedure as above, we define:
\begin{equation}
\delta({\cal A}^2)\equiv{\cal A}^2-\frac{1}{4}. \label{A2_bar}
\end{equation}
and find:

\begin{equation}
\delta({\cal A}^2)\approx\frac{1}{4}\left[
-\frac{P_{\phi}'(0)^4+2P_{\phi}'(0)^2\left(P_{\theta}'(0)+2\beta P_s'\right)^2}{\left[\left(P_{\theta}'(0)+2\beta P_s'
\right)^2+P_{\phi}'(0)^2\right]^2} \right]. \label{A2_bar_approx}
\end{equation}
By thermally averaging this, and taking only the leading term in the asymptotic expansion, we arrive at

\begin{eqnarray}
\langle\delta({\cal A}^2)\rangle
&\approx&-\frac{1}{8P_s'}\int_{-\infty}^{\infty}
dP_{\theta}'(0)dP_{\phi}'(0)|P_{\phi}'(0)|\nonumber\\ &\times&
\int_{-\infty}^{\infty}d\gamma
\frac{1+2\gamma^2}{(\gamma^2+1)^2}\nonumber\\
&\times&\frac{\exp\left\{-\frac{1}{2}\left
[P_{\theta}'(0)^2+P_{\phi}'(0)^2
\right]\right\}}{Q_{rot}}\nonumber\\&=&-\frac{3}{16P_s'}\cdot
\sqrt{\frac{\pi}{2}}, \label{Avg_classical_A2_simplified2}
\end{eqnarray}
and
\begin{equation}
\langle {\cal A}^2 \rangle \approx \frac{1}{4}-\frac{3}{16P_s'}\cdot
\sqrt{\frac{\pi}{2}}. \label{Final_average_A2}
\end{equation}
The variance can be calculated from Eqs. (\ref{Final_average_A}) and (\ref{Final_average_A2}) by using
$(\Delta {\cal A})^2=\langle{\cal A}^2 \rangle-\langle{\cal A}\rangle^2$.
Recalling the relations $P_s '=P\hbar /\sqrt{k_BTI}$ and
$J_T=\sqrt{k_B T/(hBc)}$, we have:
\begin{eqnarray}
\langle{\cal A}\rangle &\approx&
\frac{1}{2}-\frac{\sqrt{\pi}}{8}\frac{J_T}{P}\nonumber\\
(\Delta{\cal
A})^2&\approx&\frac{\sqrt{\pi}}{32}\frac{J_T}{P}.\label{Final_Results}
\end{eqnarray}

\begin{figure}[htb]
\begin{center}
\includegraphics[width=8cm]{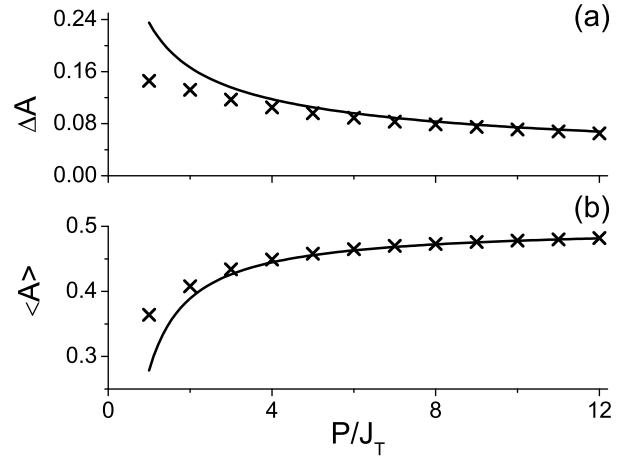}
\end{center}
\caption{Analytical asymptotic estimations of (a) $\Delta {\cal A}$ and (b)
$\langle{\cal A}\rangle$ (solid lines). The $\times$ points result from
a direct numerical calculation using Eq. (\ref{Accurate_Dist}).}
\label{AnalyticalAsymptotics}
\end{figure}

The above asymptotic expressions for $\Delta {\cal A}$ and $\langle{\cal A}\rangle$
are plotted in Figures \ref{AnalyticalAsymptotics}a and \ref{AnalyticalAsymptotics}b,
respectively (solid lines). The $\times$ points refer to the direct
numerical calculations based on the distribution function given by Eq.(\ref{Accurate_Dist}).
Although the asymptotic results (\ref{Final_Results}) are formally valid for
$J_T\geq 1$, and $P\gg J_T$, they provide a good agreement with the exact numerical
simulations already for $P/J_T =2$. Moreover, our classical asymptotic estimate for the width
of the distribution, $\Delta {\cal A}$ coincides within the $10\%$
accuracy with the exact quantum result for $P=25$ and $J_T=5$ presented above.

\section{Classical Treatment: strong deflecting field} \label{classical_treatment_strong}

In the case of a strong deflecting field, the rotational motion
of the molecules may be disturbed by the field. The Hamiltonian
of a molecule in the vertically polarized optical electric field of a constant
amplitude  $E$ is:

\begin{equation}
H=\frac{1}{2}I\left( \dot{\theta}^2+\dot{\phi}^2\sin^2\theta
\right)-\frac{1}{4} E^2 \left( \Delta
\alpha\cos^2\theta+\alpha_{\perp}\right) \label{Energy_Equation}
\end{equation}

The conjugate momenta  $P_{\phi}$ and $P_{\theta}$ are given by
Eqs. (\ref{Pphi}) and (\ref{Ptheta}), where $P_{\phi}$ is a constant of
motion for the chosen polarization. It is convenient \cite{Goldstein}
to introduce a new variable

\begin{equation}
u\equiv \cos^2\theta \label{u_constant}
\end{equation}
 that satisfies the equation

\begin{equation}
\left ( \frac{du}{dt} \right
)^2=4u\left[(1-u)\beta-\left(\frac{P_{\phi}}{I}\right)^2+(1-u)\alpha
u\right]\equiv g(u). \label{g_definition}
\end{equation}
The coefficients $\alpha , \beta $ in the polynomial $g(u)$ are given by
\begin{equation}
\beta\equiv \frac{2}{I}\left(H+\frac{1}{4}E^2\alpha_{\perp} \right)
\label{beta_constant}
\end{equation}
and
\begin{equation}
\alpha\equiv \frac{\Delta\alpha E^2}{2I}. \label{alpha_constant}
\end{equation}
Equation (\ref{g_definition}) can be immediately solved by separation of variables

\begin{equation}
dt=\frac{du}{\sqrt{g(u)}}. \label{dt_du_relation}
\end{equation}

In the case of a free rotation ($\alpha=0$), there are only two roots to
the polynomial $g(u)$: $u_2=0$ and $0 \leq u_3 \leq 1$, and $u$ performs periodic
oscillatory motion between them. When $\alpha\neq 0$, $g(u)$ generally has three roots
($u_1<u_2<u_3$), one of them is necessarily zero. For weak fields,
the middle root ($u_2$) stays at zero and the molecule performs distorted full
rotations. When increasing the field, a bifurcation  happens with the roots of
$g(u)$: the smallest root $u_1$ becomes stuck at $u=0$, and the system oscillates
in the region $0<u_2<u_3<1$ where $g(u)$ is positive. This corresponds to the so-called
pendular motion \cite{Friedrich}, when the molecular angular motion is trapped by the
external field.

Since molecules experience a time-varying amplitude of the optical field while propagating through
the deflecting beam, the total rotational energy of the system and the position of the roots $u_{1,2,3}$ are
changing with time. However, these changes are
adiabatic with respect to the rotational motion, and therefore we can
use adiabatic invariants to determine the energy of the system
\cite{Goldstein,Landau,Dugourd}. The adiabatic invariant related
with the coordinate $\theta$ is:
\begin{equation}
I_{\theta}=\oint P_{\theta}d\theta .
\label{adiabatic_invariant}
\end{equation}
It is easy to derive from Eqs.(\ref{Ptheta}), (\ref{g_definition}) and
(\ref{adiabatic_invariant}) that :
\begin{equation}
I_{\theta}=\frac{I}{4} \int_{u_2}^{u_3} \frac{\sqrt{g(u)}}{u(1-u)}du .
\label{adiabatic_invariant_modification}
\end{equation}
The energy $H$ of the molecule inside the deflecting field as a
function of the initial energy $H_0$ (before entering the field) is obtained
numerically by solving the Eq.:
\begin{equation}
I_{\theta}=I_{\theta}^0\label{numerical_E_adiabatic_invariant}
\end{equation}
where $I_{\theta}^0$ is calculated for $\alpha=0$, i.e. in the absence
of the external field.

Once the energy of the system $H$ and the polynomial $g(u)$ have
been found, the average alignment factor is simply given by:
\begin{equation}
\langle \cos^2\theta \rangle =\langle u \rangle = \frac{\int_{u_2}^{u_3}
udu/\sqrt{g(u)}}{\int_{u_2}^{u_3} du/\sqrt{g(u)}}
\label{Average_alignment_adiabatic_invariant}
\end{equation}

\begin{figure}[htb]
\begin{center}
\includegraphics[width=8cm]{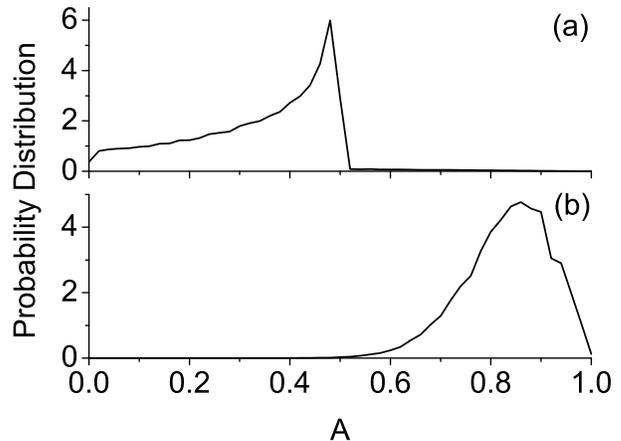}
\end{center}
\caption{Distribution of ${\cal A}$ at
the peak of the deflecting field. Two cases are considered: (a) weak deflecting field
$3\cdot 10^9 W/cm^2$ and (b) strong deflecting field
$9\cdot 10^{11} W/cm^2$. In (a) we observe the unimodal-rainbow
distribution discussed in the previous sections. In (b), the effect of the
alignment by the deflecting field is evident. }\label{Strong_Distribution_Alignment}
\end{figure}

To illustrate the performance of the procedure at real experimental conditions
\cite{Deflection_general,Barker-new}, we consider the deflection of
$CS_2$ molecules at $T=5K$ (see Fig. \ref{DeflectionScheme}), and
plot the distribution of ${\cal A}$ at the \emph{peak} of the
deflecting field. The results are given in Figs.
\ref{Strong_Distribution_Alignment}a and b, for weak ($3\cdot
10^9 W/cm^2$) and strong ($9\cdot 10^{11} W/cm^2$) deflecting fields,
respectively.

In the case of weak field, we get a unimodular rainbow distribution similar to that
derived by the various methods in the previous
sections. In the case of strong field we obtain a
rotationally-trapped distribution, corresponding to the pendular-like motion
of the molecules at the top of the deflecting pulse \cite{Barker-new}.
\\

To study the deflection  of $CS_2$ molecules by a focused laser beam, we integrate
numerically Eq.(\ref{Velocity_Deflection}) to find the deflection velocity. In the
integrand of Eq.(\ref{Velocity_Deflection}), we substitute the value of
$\langle\cos^2\theta\rangle$ calculated by Eq.(\ref{Average_alignment_adiabatic_invariant})
in every point of time. As in the
previous sections, we  assume that $x\approx v_x t$
($v_x=500m/s$) and  consider $z$ as a fixed impact parameter
($z=-4\mu m$). These assumptions are  valid even for strong
deflecting fields (that align the molecules) since the deflection
angle is still small. We consider both weak and strong deflecting fields, as in
Fig. \ref{Strong_Distribution_Alignment}, and use the values of
$\omega_0=7\mu m$ and $\tau=14ns$ (Eq. \ref{E_deflect_field}) in the
calculation of the trajectories.

\begin{figure}[htb]
\begin{center}
\includegraphics[width=8cm]{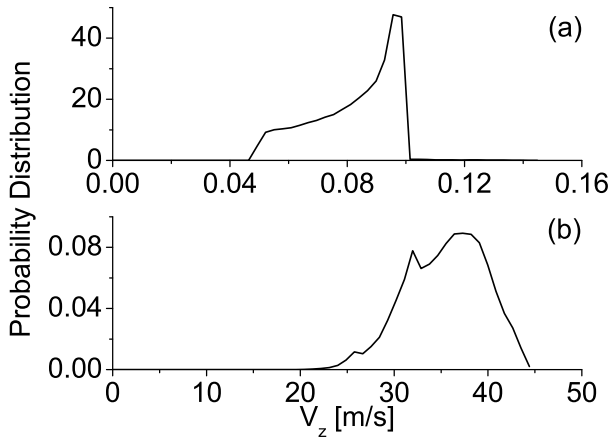}
\end{center}
\caption{Distribution of velocities (or deflecting angles)
calculated from the trajectories of the molecules subject to weak
(a) and strong (b) deflecting fields. The fields characteristics are
given in the text. }\label{Strong_Distribution_v_z}
\end{figure}

The distribution of deflected velocities for a thermal
molecular ensemble (without prealignment) is
shown in Fig. \ref{Strong_Distribution_v_z}.

In Fig. \ref{Strong_Distribution_v_z}a (weak field) we essentially
verify our assumption from the previous sections, that  the
deflection in weak fields is linear with ${\cal A}$ (Eq.
\ref{Deflection Angle}). This is  seen by observing that Fig.
\ref{Strong_Distribution_v_z}a may be indeed  obtained by a linear
transformation of the distribution $f({\cal A})$ from Fig.
\ref{Strong_Distribution_Alignment}a. In Fig.
\ref{Strong_Distribution_v_z}b (strong field) the distribution of
deflection angles (or of deflection velocities) is still quite broad. To our opinion,
this results from two different regimes of scattering that the molecules experience while
traversing the deflecting beam: weak deflection at the periphery of the beam, and deflection under
partial alignment of the molecular ensemble in the center of the beam.

\begin{figure}[htb]
\begin{center}
\includegraphics[width=8cm]{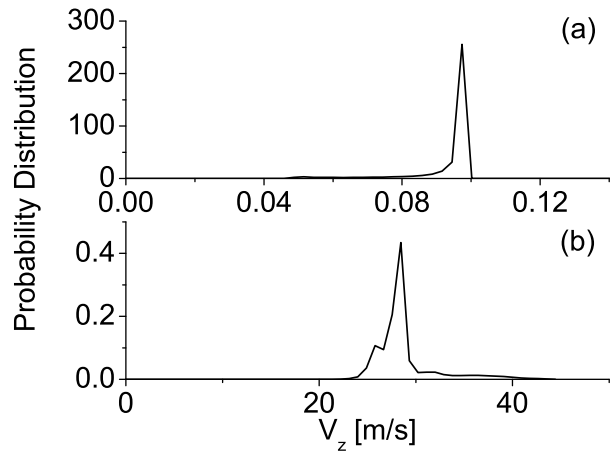}
\end{center}
\caption{Distribution of deflection velocities for prealigned
molecules ($P=25$). Case (a) is for weak deflecting field, and case
(b) is for strong deflecting field. In both cases the molecules were
prealigned parallel to the deflecting field, and this prealignment
was strong enough to ignore any aligning effect of the deflecting
field (thus reducing the broadening of the deflection, as was
explained in the previous
sections).}\label{Strong_Distribution_v_z_alignment}
\end{figure}

Finally we consider scattering of molecules prealigned in the $z$
direction with the pulses having kick strength of $P=25$.
The results are given in Figs.
\ref{Strong_Distribution_v_z_alignment}a and b for weak and strong
deflecting fields, respectively.

In the case of weak deflection (Fig.
\ref{Strong_Distribution_v_z_alignment}a), the narrow peak is observed whose
nature was already explained above. More interestingly, the distribution of the deflection angles
regained the narrow shape even in the case of strong deflection field (Fig.
\ref{Strong_Distribution_v_z_alignment}b) as a result of prealignment! In our example,
the prealignment pulse was strong enough to overcome the
rotational trapping by the deflecting field. As a result, all the
molecules performed full rotations (but not a pendular motion) despite the presence of the
strong deflecting field, and we obtain a narrow distribution as well.

\section{Discussion and Conclusions} \label{Discussion_and_conclusions}

Our results indicate that prealignment provides an effective tool
for controlling the deflection of rotating molecules, and it may be
used for increasing the brightness of the scattered molecular beam.
This increase was shown both for weak and strong deflecting fields.
This might be important for nano-fabrication schemes based on the
molecular optics approach \cite{Seideman}. Moreover, molecular
deflection by non-resonant optical dipole force is considered as a
promising route to separation of molecular mixtures (for a recent
review, see \cite{Chinese}). Narrowing  the distribution of the
scattering angles may substantially increase the efficiency of
separation of multi-component beams, especially when the
prealignment is applied selectively to certain molecular species,
such as isotopes \cite{isotopes}, or nuclear spin isomers
\cite{Fauchet,isomers}. More complicated techniques for pre-shaping
the molecular angular distribution may be considered, such as
confining molecular rotation to a certain plane by using the
"optical molecular centrifuge" approach \cite{Centrifuge},
double-pulse ignited "molecular propeller" \cite{conf1,conf2,NJP,York,Ohshima},
or planar alignment by perpendicularly polarized laser pulses
\cite{alternation}. In this case, a narrow angular peak is expected
in molecular scattering, whose position is controllable by
inclination of the plane of rotation with respect to the deflecting
field \cite{Floss}. Laser prealignment may be used to manipulate molecular
deflection by inhomogeneous \emph{static} fields as well \cite{static} (for recent
exciting experiments on \emph{post-alignment} of molecules scattered
by static electric fields see \cite{post}). In particular, one may
affect molecular motion in relatively weak fields that are
insufficient to modify rotational states by themselves.  Moreover,
the same mechanisms may prove efficient for controlling inelastic
molecular scattering off metalic/dielectric surfaces. These and
other aspects of the present problem are subjects of an ongoing
investigation.

This research is made possible in part by the historic generosity of
the Harold Perlman Family. IA is an incumbent of the Patricia Elman
Bildner Professorial Chair. We thank Yuri Khodorkovsky for providing us the results of his related time-dependent quantum calculations.

\bibliographystyle{phaip}

\end{document}